\documentclass[sigchi,screen]{acmart}
\AtBeginDocument{%
  \providecommand\BibTeX{{%
    \normalfont B\kern-0.5em{\scshape i\kern-0.25em b}\kern-0.8em\TeX}}}

\setcopyright{acmcopyright}
\copyrightyear{2023}
\acmYear{2023}
\acmDOI{XXXXXXX.XXXXXXX}

\acmConference[CHI '23]{Proceedings of the 2023 CHI Conference on Human Factors in Computing Systems}{April 23--28, 2023}{Hamburg, Germany}
\acmBooktitle{Proceedings of the 2023 CHI Conference on Human Factors in Computing Systems (CHI '23), April 23--28, 2023, Hamburg, Germany} 
\acmPrice{15.00}
\acmISBN{978-1-4503-XXXX-X/18/06}

\begin{document}

\title[Empathic Metaverse]{The Empathic Metaverse: An Assistive Bioresponsive Platform For Emotional Experience Sharing}

\author{Yun Suen Pai}
\email{pai@kmd.keio.ac.jp}
\affiliation{%
  \institution{Keio University Graduate School of Media Design}
  \country{Japan}
}

\author{Mark Armstrong}
\email{mark@keio.jp}
\affiliation{%
  \institution{Keio University Graduate School of Media Design}
  \country{Japan}}

\author{Kinga Skierś}
\email{kinga.skiers@gmail.com}
\affiliation{%
  \institution{Keio University Graduate School of Media Design}
  \country{Japan}
}

\author{Anish Kundu}
\email{anish@kmd.keio.ac.jp}
\affiliation{%
 \institution{Keio University Graduate School of Media Design}
 \country{Japan}}

\author{Danyang Peng}
\email{pengdanyang@keio.jp}
\affiliation{%
  \institution{Keio University Graduate School of Media Design}
  \country{Japan}}

\author{Yixin Wang}
\email{ywang09@keio.jp}
\affiliation{%
  \institution{Keio University Graduate School of Media Design}
  \country{Japan}}

\author{Tamil Selvan Gunasekaran}
\email{themastergts007@gmail.com}
\affiliation{%
  \institution{University of Auckland}
  \country{New Zealand}}

\author{Chi-Lan Yang}
\email{chilan.yang@cyber.t.u-tokyo.ac.jp}
\affiliation{%
  \institution{The University of Tokyo}
  \country{Japan}}

\author{Kouta Minamizawa}
\email{kouta@kmd.keio.ac.jp}
\affiliation{%
  \institution{Keio University Graduate School of Media Design}
  \country{Japan}}

\renewcommand{\shortauthors}{Pai, et al.}

\raggedbottom
\begin{abstract}
The Metaverse is poised to be a future platform that redefines what it means to communicate, socialize, and interact with each other. Yet, it is important for us to consider avoiding the pitfalls of social media platforms we use today; cyberbullying, lack of transparency and an overall false mental model of society.
In this paper, we propose the Empathic Metaverse, a virtual platform that prioritizes emotional sharing for assistance. It aims to cultivate prosocial behaviour, either egoistically or altruistically, so that our future society can better feel for each other and assist one another. To achieve this, we propose the platform to be bioresponsive; it reacts and adapts to an individual's physiological and cognitive state and reflects this via carefully designed avatars, environments, and interactions. We explore this concept in terms of three research directions: bioresponsive avatars, mediated communications and assistive tools.
\end{abstract}



\begin{teaserfigure}
  \includegraphics[width=\textwidth]{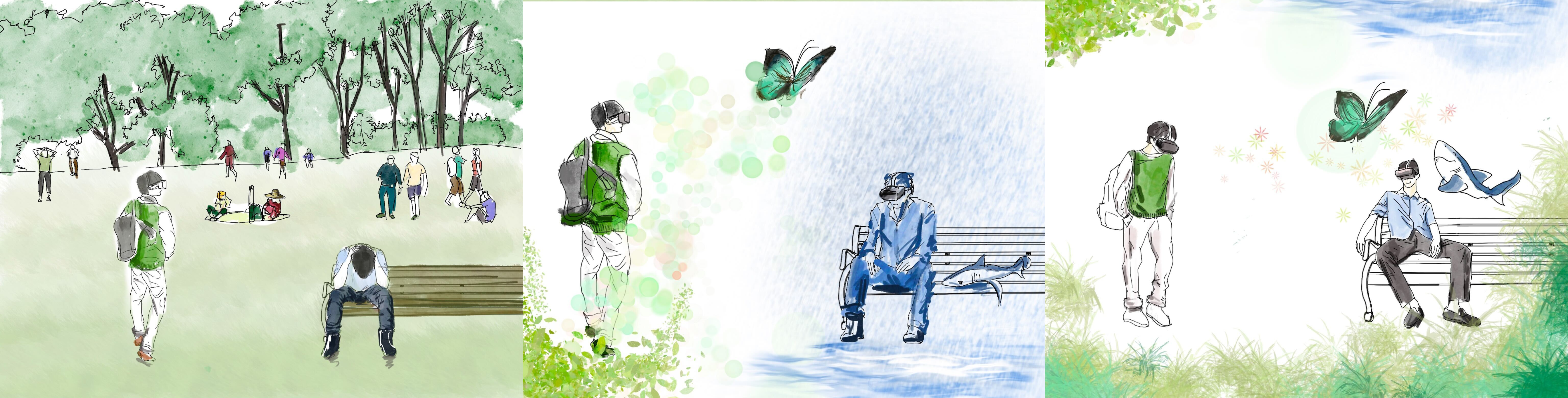}
  \caption{A conceptual interaction in the Empathic Metaverse: A user can visually and explicitly feel and understand another more than what is possible in physical space, triggering an interaction that can assist the other's mental state}
  \label{fig:teaser}
\end{teaserfigure}

\received{20 February 2007}
\received[revised]{12 March 2009}
\received[accepted]{5 June 2009}

\maketitle

\section{Introduction}
Social connections between individuals are a popular and multi-faceted research topic. Technology that facilitates remote social connection these days are social media platforms with the likes of Facebook\footnote{\url{https://www.facebook.com/}}, Twitter\footnote{\url{https://twitter.com/}}, Instagram\footnote{\url{https://www.instagram.com/}}, and so on. These platforms allow their users to share moments and stories of their lives easier.
Yet, as beneficial as those platforms are, it creates a false mental model of society. Users of these platforms tend to only share and emphasize their accomplishments, creating an illusion of a perfect lifestyle \cite{steers2014seeing}. 
By doing so, it not only encourages them to hide their emotional state, but also displays to others a false reality that potentially leads to poorer mental well-being \cite{o2018social}. Being able to express emotions is an important social tool for humans, and care needs to be taken on not just providing a platform to facilitate this properly, but also ensuring it is inclusive and accessible to diverse users.

Our abilities to express and share our emotions as human beings are rather subtle by nature. Implicit expressions, like the twitch of the eyebrow, a curve of the lips, and other forms of body language are utilized for us to convey our emotions in a more indirect manner and we expect the receiver to be able to interpret them correctly.
Affective computing \cite{picard2000affective} is a research area that aims to detect, interpret, and simulate emotions from these cues. However, the subtleness in how we express and make ourselves understood by others emotionally makes it a consistent challenge.
What if we can understand and feel the emotions and mental state of another clearly simply by just looking at them? What if there was a visual representation that is so clear and distinct, allowing us to immediately determine whether a person is sad, happy, stressed, or calm, allowing us to empathize with them?
The advent of the Metaverse can facilitate new modalities of emotional connection previously not possible using conventional social media. For example, body language can now be understood when communicating and socializing with remote users \cite{smith2018communication}.
There were also several related works regarding carrying over facial expressions into the Metaverse to improve empathy \cite{yalccin2020empathy, lisetti2013can}, though these methods focus on importing conventional methods of emotional expression into the Metaverse. We argue that, as a future platform, we should leverage the tools provided within the Metaverse to promote a form of empathy that is easier, better, and more socially connected than in real life. To achieve this, we require a new mental model of emotion as a sensory stimulus.

We propose the Empathic Metaverse, a virtual platform that leverages individuals' physiological states to derive their emotion and cognition. This information is then reflected in the Metaverse in the form of bioresponsive avatars, environments and interactions. We define bioresponsive as a system within the Metaverse that reacts and adapts to physiological states, making them more perceivable to the self and other individuals beyond what physical reality is capable of. 
It is related to biofeedback systems \cite{yu2018biofeedback} that increase awareness of our physiological states and are commonly used in therapy. Song et al. \cite{song2017framework} proposed a framework for bio-responsive VR that describes how physiological data can be mapped to VR in terms of the movement and gestures.
It is also closely related to empathic computing and mixed reality \cite{piumsomboon2017empathic} which enables us to share feelings and interactions. The authors explored how physiological-driven modalities can assist in remote collaboration. For the Empathic Metaverse, we plan to build on these works by defining a taxonomy of visual stimuli as a form of a new mental model for empathy. We explore how an individual's emotional and cognitive state can be seen, understood, and subsequently felt through the appearance of their avatar, how social communication is mediated, and how it can ultimately transform into assistive tools. We also discuss how we can learn to accept this new model of perceiving empathy, how empathy can translate to actionable assistance, and how we should consider this in terms of limitations and ethical concerns.
\section{Mental Model of The Empathic Metaverse}
The Empathic Metaverse is a novel concept of perceiving empathy for our Metaverse future. It proposes a new mental model of empathy derived from our emotional interpretation of virtual avatars, environments, and our interactions with them. We break this down into three directions: bioresponsive avatars, mediated communications and assistive tools.

\subsection{Bioresponsive Avatars}
Bioresponsive avatars refer to an avatar system where each user's physiological state is measured and classified into their respective emotional and cognitive state. Then, these signals are used to drive an avatar's representation in the Metaverse. We use physiology because it is an involuntary and natural reaction to a context as a human being.
In a recent work by Liu et al. \cite{liu2017can}, the authors introduced expressive biosignals, where they explored visualizing brain signals in terms of graphs, sliders, swirls, colors, lights and emoji. 
Ferstl et al. \cite{ferstl2016trust} found that design factors like head shape, eye shape, and eye size can influence our perception of aggressiveness, dominance, trustworthiness, appeal and eeriness.
Lee et al. \cite{lee2022understanding} explored how physiological signals can be visualized around one's avatar and they found that skeuomorphic visualizations for physiological signals can be inferred as different arousal levels. However, these works have yet to present an emotional mapping of visualization elements within the Metaverse.
A close related work, Emotional Beasts \cite{bernal2017emotional}, also aims to create an expressible and abstract avatar. They changed particle size, density, brightness and colour for different emotional expressions, yet the mapping was not clearly defined and not applied to interactions between users.

\subsection{Mediated Communications}
\begin{figure}
  \includegraphics[width=\columnwidth]{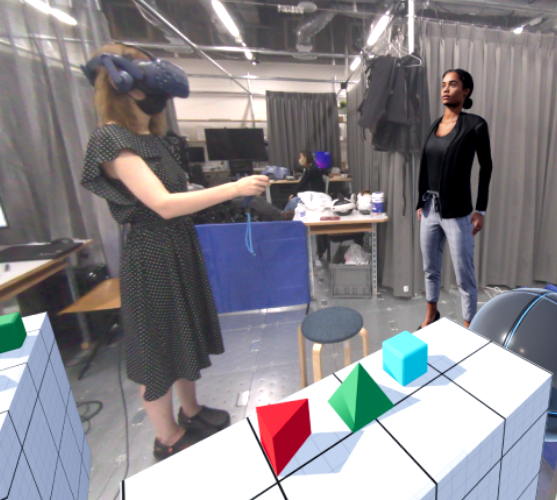}
  \caption{A study to determine how avatar gender expression affects social interaction in virtual reality \cite{kundu2022investigating}.}
  \label{fig:lgbtech}
\end{figure}

Mediated communications is a social platform within the Empathic Metaverse where communications are mediated, enhanced and improved.
Some related works include Animo \cite{liu2019animo} which uses a geometrical avatar to facilitate interaction, whereas Significant Otter \cite{liu2021significant} improves on this by including a pair of well-animated otters. Both these systems are driven by physiological signals and have proven to improve presence and communication. 
Yet, we intend to explore how this can be brought forward to the Metaverse and what kind of mediation can be adopted in an immersive space. For example, an investigation on collective emotional state and how this can be leveraged to drive an avatar to enhance group communication based on its bioresponsive properties. In an immersive yet few social cues, such as facial expressions, environment, it is important to explore how biosignals can be designed and presented to mediate social interaction in metaverse.  
In a social scenario, we also need to consider how to accommodate a diverse group of users, regardless of their physical and mental activities, personal preferences, orientation and background. As an initial investigation, we explored how gay cisgender men perceive and interact with avatars of varying gender expression and found that the emotion of female-looking avatars were more well received by the receivers \cite{kundu2022investigating}, as shown in Figure \ref{fig:lgbtech}.

\subsection{Assistive Tools}
Though understanding and feeling for others is important, it is equally important to provide assistance.
Assistive tools refer to how the platform can translate empathy to a tangible or actionable event that can assist users in terms of their physical or mental well-being. 
An initial approach is Transcendental Avatar \cite{skiers2022transcendental}, a system that provides users with an enhanced emotional perception of the self. We carefully designed an abstract representation of the mirrored self that reflects our cognitive state predicted via electrodermal activity and heart rate, and use this to regulate stress levels.
Previous studies have shown that virtual environments can be used to teach arousal or stimulate empathy \cite{paananen2022digital}. One example of this is the use of virtual environments in the training of medical and educational staff. By assuming the role of victims of abuse, for example, staff can gain a deeper understanding of violent behavior and the perspectives of abused individuals through perspective-taking approaches. Paananen et al. \cite{paananen2022digital} have also explored the minimum level of user bandwidth required to elicit empathetic experiences in virtual environments. Their research highlights that few studies have explored the accuracy of user experiences in such virtual empathy scenarios, and there is a dearth of knowledge regarding the degree of fidelity that virtual experiences bear to the real-life experiences they simulate. Furthermore, as existing works employ distinct virtual settings and empathetic goals, there are significant opportunities for future research to explore the connection between the design of fictional scenarios and the process of empathizing with users' personal experiences
Another work, It's Me \cite{wang2022s}, is a journaling tool that allows users to log their daily experiences not just through text and words, but through an embodied avatar interaction and can be replayed for cognitive self-regulation. 
Our next step is to target users with disabilities, understand how the representations in the Empathic Metaverse can assist them, co-design the bioresponsive platform with them, and deploy it in longitudinal studies for evaluation.

\begin{figure}
  \includegraphics[width=\columnwidth]{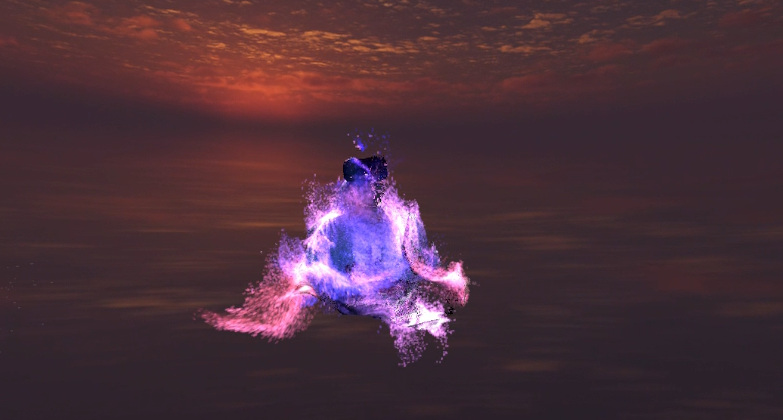}
  \caption{A demonstration on how abstract avatar self-perception can be used to reduce stress \cite{skiers2022transcendental}.}
  \label{fig:transcendental}
\end{figure}

\section{Initial Workshop and Findings}
The goal of the initial workshop is to derive a basic design taxonomy regarding how bioresponsive elements can be understood based on emotion models.
We conducted a preliminary workshop with 12 participants (9 females, mean:23.75, SD:1.64) regarding how avatar design and appearance affect our emotional and cognitive state. The workshop  was conducted in groups of four via Zoom\footnote{\url{https://zoom.us/}} and Miro\footnote{\url{https://miro.com/app/}}. It was initiated with simple questions about how they would like to have their avatar appear in a virtual world. They were then given 5 minutes to illustrate this using their desired sketching paradigms. After that, each participant was given 5 minutes to explain the logic behind their design. This procedure was repeated four more times, where we defined an emotion of the user for each design (happy, sad, calm, and stressed) and how they wished to have it reflected on their avatar. Finally, a group discussion was held where users expressed their opinions and provided feedback regarding each other's design.

Based on the collected results, we generated a taxonomy of the mapping of avatar and environment design according to the Valence-Arousal model first proposed by Russell \cite{russell1980circumplex}. Colours overall play an important role in portraying emotional states, evident from previous work findings as well \cite{liu2017can}. When asked about an avatar appearance and form when feeling happy (positive valence and high arousal), blueish colour with friendly anthropomorphic avatars conveys this well. Particles and surround effects can also look fiery and glittery. In contrast, a sad emotion (negative valence and low arousal) can be reflected with muted tones and objects and avatars that are more elliptical. Elements of water and shadows can also be used for a sadder atmosphere. For a stressful emotion (negative valence with high arousal), angular shapes with broken effects and animations reflect this well. Avatars can also appear to look sweaty and steamy. Lastly, for calm emotion (positive valence with low arousal), a greener or cooler space paired with low-key animations is helpful. Round and soft textures also bring an element of comfort. 

\begin{figure}
  \includegraphics[width=\columnwidth]{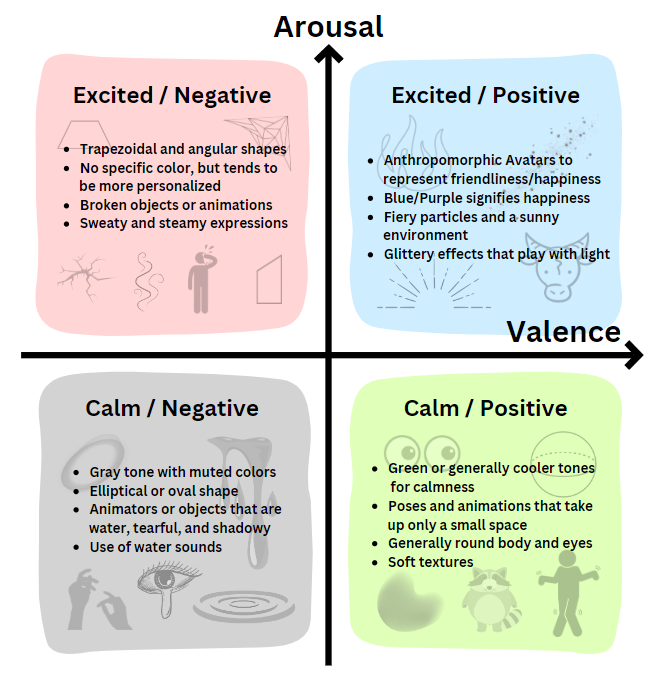}
  \caption{A brief taxonomy on the design of Empathic Metaverse that covers the visual representation for avatar, environment and interaction based on the Valence-Arousal Model \cite{russell1980circumplex}.}
  \label{fig:taxonomy}
\end{figure}

\section{Design Considerations}
As with the design of any new platform, we should take the context and scenario into account before deciding on an appropriate visual stimulus. Our preliminary findings are meant for generic social environments in the Metaverse, and we have yet to collect data regarding how the bioresponsive platform should adapt to various contexts, such as workplace environments, gaming activities, and so on.
Furthermore, a platform that can potentially improve a user's ability to empathize may sound beneficial, yet, care must be taken to ensure it remains a safe and ethical platform. Empathy should be regulated carefully; otherwise, it may potentially lead to other forms of disorders instead. For example, hyper-empathy syndrome is a mental health issue that causes dysfunctional empathic emotional overreaction \cite{green2007cognitive, shamay2009neuropsychological}. Exposure to negative emotional events may cause the patient to feel an exaggerated and disproportionate amount of sadness, grief, cognitive load, and other forms of physiological reactions.
A potential limitation of the Empathic Metaverse platform is also the involvement, or lack thereof, of those not within the Metaverse \cite{UIST20_HMDlight}. As our method proposes a new concept of perceiving empathy made possible only by augmenting reality, we should also ensure any additional knowledge or insight can still affect interactions in physical space.
Lastly, we need to always consider about privacy, especially for a platform that integrates physiological and emotional feedback. Previous work has shown that, although people want to perceive emotions better, they are not as willing to share them \cite{lee2022understanding, hassib2016investigating}. Proper explanations, rules, and protocols regarding the usage of personal data and information should be designed before a platform like this can be deployed to the public.
\section{Conclusion and Future Works}
Our vision of an inclusive and accessible Metaverse is a new mental model of empathy; the ability to visually perceive and lead to a better understanding of an individual's emotional and cognitive state. We further break this down into three research directions, which are bioresponsive avatars, mediated communications, and assistive tools.
We believe that, with the Empathic Metaverse platform, it can foster prosocial behaviour change so that methods and solutions for the future of the inclusive and accessible Metaverse can be initiated. In parallel, we will also expand our investigation towards other forms of sensory stimuli and how they can be mapped into our extended taxonomy in the Metaverse in the future.

\section{Acknowledgments}

This work was supported by the JST Moonshot R\&D Program "Cybernetic Being" Project (Grant Number JPMJMS2013).

  \bibliographystyle{ACM-Reference-Format}
  \bibliography{0-References}

\end{document}